\documentstyle[seceq]{ptptex}

\def\tr{\mathop{\rm tr}\nolimits}

\makeatletter
\@addtoreset{equation}{section}
\makeatother

\newcommand{\VEV}[1]{\left\langle #1 \right\rangle}

\newcommand{\Mp}{M_P}

\notypesetlogo  

\title{
$E_6$ Unification with Bi-Large Neutrino Mixing
}

\author{
Masako {\sc Bando}\footnote{
E-mail: bando@aichi-u.ac.jp}
and 
Nobuhiro {\sc Maekawa}\footnote{
E-mail: maekawa@gauge.scphys.kyoto-u.ac.jp}
}

\inst{
Physics Division, Aichi University, Aichi 470-0296, Japan \\
$^{**}$Department of Physics, Kyoto University, Kyoto 606-8502, Japan
}

\recdate{
}

\abst{
The scenario of $SO(10)$ grand unified theory (GUT)
proposed by one of the authors is extended to $E_6$ unification.
This gives realistic quark and lepton mass matrices. In the 
neutrino sector,
the model reproduces the large mixing angle (LMA) MSW solution 
as well as the large mixing angle for the atmospheric neutrino anomaly.
In this model, the right-handed down quark and the left-handed lepton of the
first and second generations belong to a single multiplet ${\bf 27}$.
This causes natural suppression of
the flavour changing neutral current(FCNC).
}

\begin{document}

\maketitle

\section{Introduction}
There are strong reasons to believe in the validity of grand unified theories 
(GUT)\cite{georgi}, 
in which the quarks and leptons are unified
in several multiplets in a simple gauge group.
They explain various matters that cannot be understood within the
standard model: the `miracle' of anomaly cancellation between quarks
and leptons, the hierarchy
of gauge couplings, charge quantization, etc.
The three gauge groups in the
standard model are unified into a simple gauge group at a GUT scale, which
is considered to be just below the Planck scale.
On the other hand, the GUT scale destabilizes the weak scale. 
One of the most promising
ways to avoid this problem is to introduce supersymmetry (SUSY).
However, it is not easy to obtain a realistic SUSY 
GUT.\cite{SUSYGUT}
First, it is difficult to obtain realistic
fermion mass matrices.
Also unification of quarks and leptons puts strong
constraints
on the Yukawa couplings.
Finally, one of the most difficult obstacles is
the ``doublet-triplet (DT) splitting problem."

There have been several attempts to solve the DT splitting 
problem.\cite{DTsplitting,DW} 
Among them, the Dimopoulos-Wilczek mechanism is
 a promising way to realize
DT splitting in the $SO(10)$ SUSY 
GUT.\cite{DW,BarrRaby,Chako,complicate}

Concerning the fermion masses,
recent progress in neutrino experiments\cite{SK} 
provides important information on  family structure.
There are several impressing 
works\cite{Sato,Nomura,Bando,Barr,Shafi} in which 
the large neutrino mixing angles are realized within GUT framework.
It is now natural to examine
$SO(10)$ and higher gauge groups because they allow for every quark 
and lepton, including the right-handed neutrino, to be unified in 
a single multiplet, which is
important in addressing neutrino masses.

Recently, one of the authors (N. M.)\cite{maekawa} proposed
 a scenario of $SO(10)$ grand unified theory (GUT)
 with anomalous $U(1)_A$ gauge symmetry,
 which has the following interesting features:
\begin{enumerate}
\item
The doublet-triplet (DT) splitting is realized
using the Dimopoulos-Wilczek mechanism.
\item
Proton decay via the dimension-five operator is suppressed.
\item
Realistic quark and lepton mass matrices can be obtained in a simple
way. In particular, in the neutrino sector, bi-large neutrino 
mixing is realized.
\item
The symmetry breaking scales are determined by the anomalous $U(1)_A$
charges.
\item
The mass spectrum of the super heavy particles is fixed by the anomalous
$U(1)_A$ charges.
\end{enumerate}

As a consequence of the above features, the fact that the GUT scale is
smaller than the Planck scale
leads to modification of the undesirable GUT relation
between the Yukawa couplings 
$y_\mu=y_s$ (and also $y_e=y_d$)
while preserving the relation $y_\tau=y_b$. Moreover, it is remarkable 
that the interaction
is generic; all the interactions that are allowed 
by the symmetry are taken into account. Therefore, once we 
fix the field contents with their quantum numbers, all the 
interactions are determined, except coefficients
of order 1.

The anomalous $U(1)_A$ gauge symmetry,\cite{U(1)}
whose anomaly is cancelled by the Green-Schwarz 
mechanism,\cite{GS}
 plays an essential role
in explaining the DT splitting mechanism at the unification
scale as well as in reproducing Yukawa 
hierarchies.\cite{Ibanez,Ramond,Dreiner}
Also, bi-large neutrino mixing is naturally
obtained by choosing the ${\bf 10}$ representation with an appropriate
$U(1)_A$ charge, in addition to the three family ${\bf 16}$ representations.
This anomalous $U(1)_A$  is a powerful tool not only to reproduce
DT splitting but also to determine the GUT breaking scales.

This paper aims to show further that
the above $SO(10)$ model is naturally extended to $E_6$ GUT,
in which the additional field ${\bf 10}$ of $SO(10)$ is
included in a chiral multiplet ${\bf 27}$ of $E_6$.
In order to realize this scenario, it is important to introduce the
concept of ``twisting family structure" in the $E_6$ unified 
model.\cite{Bando}

Under the $SO(10)$ group, we know that
 {\bf10} and ${\bf \overline 5}$ of $SU(5)$ are combined into  
 {\bf16} of $SO(10)$.
Usually, each family belongs to {\bf16}.
In this framework, however, it is not easy to reproduce
 the large Maki-Nakagawa-Sakata (MNS)\cite{MNS} mixing and
 small Cabbibo-Kobayashi-Maskawa (CKM)\cite{CKM} mixing.
 A promising way to reproduce this is to introduce other multiplets,
 ${\bf 10}$ of $SO(10)$, in addition to the usual $3\times 
 {\bf 16}$ multiplets.\cite{Nomura}
Since ${\bf 10}$ of $SO(10)$ is decomposed into ${\bf 5}({\bf 10})$ 
and
${\bf \overline 5}({\bf 10})$ of $SU(5)$, one of the fields
${\bf \overline 5}({\bf 16})$ can be replaced by this
${\bf \overline 5}({\bf 10})$. Such a replacement is essential to 
reproduce
large MNS mixing, preserving small CKM mixing.
In the case of $E_6$, ${\bf 16}$ and ${\bf 10}$ of $SO(10)$ are 
naturally
included in a single multiplet {\bf 27} of $E_6$.
The $E_6$ model automatically prepares such a replacement,
as we see in the next section. We call the mechanism responsible for this 
the ``twisting mechanism."
This gives us a strong motivation to examine $E_6$ unification.

It is interesting that the above desired scenario in $SO(10)$
unification can be extended to $E_6$ unification while keeping the
desirable features of $SO(10)$ unification.
In this paper we focus on the extension of the matter sector to
$E_6$ unification, leaving discussion of DT splitting to a
separate paper.\cite{BKM} The extension of DT splitting 
to $E_6$ unification is non-trivial.
Moreover, we show that the condition for the suppression of
the flavour changing neutral current (FCNC) is automatically 
satisfied. This is essentially caused by the twisting 
mechanism and the unification of the matter fields into a single 
multiplet ${\bf 27}$, which guarantees that 
${\bf \overline 5}({\bf 16})$ and ${\bf \overline 5}({\bf 10})$
have the same anomalous $U(1)_A$ charge.
Then it can happen that the charge of the first
generation of ${\bf \overline 5}$ becomes equivalent to that of 
the second generation of ${\bf \overline 5}$. This weakens the 
severe constraint resulting from the
FCNC. It is interesting that the selection of the anomalous
$U(1)_A$ charge to realize bi-large neutrino mixing angles automatically
causes the above mensioned FCNC suppression.

In section 2, we briefly review the twisting mechanism 
and classify the patterns of the massless modes of the ${\bf \bar 5}$
fields.
In section 3 and the Appendix, we discuss how the SUSY vacua are
determined in the  anomalous $U(1)_A$ framework.
In section 4 and 5, we study the realistic
quark and lepton mass matrices in $E_6$ unification,
where in the neutrino sector, bi-large neutrino mixing angles are
naturally realized.
In section 6, we examine the effect of SUSY breaking.
Specifically, we are able to automatically obtain a condition for 
suppression of flavour changing neutral current process 
($K^0\bar K^0$ mixing) in $E_6$ unification.

\section{Twisting in $E_6$ unification}
Let us first  recall the twisting mechanism,
which has been proposed by one of the authors (M. B.).
\cite{Bando}
The twisting family structure arising through this mechanism is peculiar to
the $E_6$ unification model, and here we explain how it arises.
In the case of $E_6$, ${\bf 16}$ and ${\bf 10}$ of $SO(10)$ are naturally
included in a single multiplet {\bf 27} of $E_6$.
The fundamental representation of $E_6$ contains {\bf16} and
{\bf10} of $SO(10)$ automatically:
Under $E_6\supset SO(10)\supset SU(5)$, we have
\begin{equation}
{\bf27} \rightarrow \underbrace{[( {\bf 16,10}) +({\bf 16,\bar 5})
+({\bf 16,1})]}_{\bf 16}
+\underbrace{[({\bf 10,\bar 5})+({\bf 10,5})]}_{\bf 10}
+ \underbrace{[({\bf 1,1})]}_{{\bf 1}},
\end{equation}
where the representations of $SO(10)$ and $SU(5)$ are explicitly denoted.
As we have already seen, the $E_6$ model naturally possesses the freedom
to replace matter fields $({\bf 16,\overline 5})$ by
$({\bf 10,\overline 5})$.
So here let us explain how the twisting family structure arises
in the $E_6$ unification.
In order to do this, it is enough to introduce the following
Higgs fields,\footnote{
Note that the
additional Higgs fields $\bar \Phi ({\bf \overline{27}})$ and
$\bar C({\bf \overline{27}})$ are required to satisfy
the $D$-flatness condition of $E_6$ gauge theory, and an adjoint
field $A({\bf 78})$ is required to break the GUT gauge group into the 
standard gauge group. In order to realize doublet-triplet
splitting, the actual breaking pattern must be
$E_6\rightarrow SO(10) \rightarrow 
SU(3)_C\times SU(2)_L\times SU(2)_R\times U(1)_{B-L} 
\rightarrow SU(3)_C\times SU(2)_L \times U(1)_Y$.\cite{maekawa,BKM}
} 
which are necessary to determine the mass matrices
of matter multiplets $\Psi_i({\bf 27})$, whose  $U(1)_A$ charges 
are denoted as  $\psi_i$\footnote{
We assume that $\psi_1>\psi_2>\psi_3$}($i=1,2,3$):
\begin{enumerate}
\item A Higgs field that breaks $E_6$ into $SO(10)$:
$\Phi({\bf 27})$ ($\VEV{\Phi({\bf 1,1})}=v$).
\item A Higgs field that breaks $SO(10)$ into $SU(5)$:
$C({\bf 27})$ ( $\VEV{C({\bf 16,1})}=v'$).
\item A Higgs field  that includes the Higgs doublets: $H({\bf 27})$.
\end{enumerate}
Throughout this paper we denote all the superfields
with
uppercase letters and their anomalous $U(1)_A$ charges with the
corresponding
lowercase letters.
Assigning  negative R-parity to the ordinary matter $\Psi_i({\bf 27})$,
as usual, and using a field $\Theta$ with charge $-1$,
the $U(1)_A$ invariant superpotential for low energy Yukawa terms becomes 
\begin{equation}
W_Y=\left(\frac{\Theta}{\Mp}\right)^{\psi_i+\psi_j+h}\Psi_i\Psi_jH,
\end{equation}
where we omit coefficients of order 1, and
for the above we assume that $\psi_i+\psi_j+h\geq 0$
for each $i,j$ pair so that there appears no SUSY zero.\footnote{ 
Note that if the total charge
of an operator is negative, the $U(1)_A$ invariance forbids 
the existence of operators in the action,
 since the field $\Theta$ with negative 
charge cannot compensate for the negative total charge of the operator 
(the SUSY zero mechanism).}
After obtaining non-zero VEV $\VEV{\Theta}=\lambda\Mp$ 
($\lambda\sim 0.2$) through the $D$-flatness condition of the anomalous 
$U(1)_A$ gauge symmetry, a hierarchical
structure of the Yukawa couplings is realized.
Since we need $3\times[{\bf 10+\bar 5
+1}]$  in $SU(5)$ representations for three families, among the above 
three {\bf 27} fields, three pairs of $({\bf 5,\bar 5})$ must become 
heavy.\footnote{The possible right-handed neutrino modes
 $\Psi_i({\bf 16,1})$ and 
$\Psi_i({\bf 1,1})$ also acquire large masses, 
but here we concentrate on the family structure of ${\bf
 \bar 5}$.
}
Indeed, the Higgs fields $\Phi$ and $C$ can yield such masses: 
 The superpotentials, which give large masses for
 $({\bf 5},{\bf \bar 5})$ pairs, are
\begin{equation}
W=\left(\frac{\Theta}{\Mp}\right)^{\psi_i+\psi_j+c}\Psi_i\Psi_jC
  + \left(\frac{\Theta}{\Mp}\right)^{\psi_i+\psi_j+\phi}\Psi_i\Psi_j\Phi,
\end{equation}
which we analyse here to see how ${\bf \bar 5}$ fields acquire 
large masses.
The VEV $\VEV{\Phi({\bf 1,1})}=v$ gives the $3\times 3$ mass matrix  of
$\Psi_i({\bf 10,5})\Psi_j({\bf 10,\bar 5})$ pairs, 
\begin{equation}
(M_{ij})=\bordermatrix{
&\Psi_1({\bf 10, \bar 5})&\Psi_2({\bf 10, \bar5})
&\Psi_3({\bf 10,\bar 5})\cr
\Psi_1({\bf 10,5})
&\lambda^{2\psi_1}&\lambda^{\psi_1+\psi_2}
&\lambda^{\psi_1+\psi_3}  \cr
\Psi_2({\bf 10,5})& \lambda^{\psi_1+\psi_2}   &  \lambda^{2\psi_2}
&\lambda^{\psi_2+\psi_3} \cr
\Psi_3({\bf 10,5})
 &\lambda^{\psi_1+\psi_3} & \lambda^{\psi_2+\psi_3}
 & \lambda^{2\psi_3}  \cr}
 \lambda^{\phi}v,
\end{equation}
while the VEV $\VEV{C({\bf 16,1})}=v'$ gives the mass terms
of $\Psi_i({\bf 16,\overline 5})$ and $\Psi_j({\bf 10,5})$, 
\begin{equation}
(M'_{ij})=
\bordermatrix{
&\Psi_1({\bf 16,\bar 5})&\Psi_2({\bf 16,\bar 5})
&\Psi_3({\bf 16,\bar5})\cr
\Psi_1({\bf 10,5})&\lambda^{2\psi_1}&\lambda^{\psi_1+\psi_2}
&\lambda^{\psi_1+\psi_3} \cr
\Psi_2({\bf 10,5})&\lambda^{\psi_1+\psi_2}&\lambda^{2\psi_2}
&\lambda^{\psi_2+\psi_3} \cr
\Psi_3({\bf 10,5})
&\lambda^{\psi_1+\psi_3}&\lambda^{\psi_2+\psi_3}
&\lambda^{2\psi_3} \cr}
\lambda^{c}v'.
\end{equation}
Then, the full mass matrix is 
\begin{equation}
\bordermatrix{
&\Psi_1({\bf 16,\bar 5})&\Psi_2({\bf 16,\bar 5})
&\Psi_3({\bf 16,\bar5})
&\Psi_1({\bf 10, \bar 5})&\Psi_2({\bf 10, \bar5})
&\Psi_3({\bf 10,\bar 5})\cr
\Psi_1({\bf 10,5})&\lambda^{2\psi_1+r}&\lambda^{\psi_1+\psi_2+r}
&\lambda^{\psi_1+\psi_3+r}
&\lambda^{2\psi_1}&\lambda^{\psi_1+\psi_2}
&\lambda^{\psi_1+\psi_3}  \cr
\Psi_2({\bf 10,5})&\lambda^{\psi_1+\psi_2+r}&\lambda^{2\psi_2+r}
&\lambda^{\psi_2+\psi_3+r}
& \lambda^{\psi_1+\psi_2}   &  \lambda^{2\psi_2}
&\lambda^{\psi_2+\psi_3} \cr
\Psi_3({\bf 10,5}) &\lambda^{\psi_1+\psi_3+r}&\lambda^{\psi_2+\psi_3+r}
&\lambda^{2\psi_3+r}
 &\lambda^{\psi_1+\psi_3} & \lambda^{\psi_2+\psi_3}
 & \lambda^{2\psi_3}  \cr} \lambda^{\phi}v ,
 \label{full}
\end{equation}
where we have defined  the parameter $r$ as 
\begin{equation}
\lambda^r\equiv \frac{\lambda^cv'}{\lambda^\phi v},
\end{equation}
which we use frequently in the following discussion. 
Note that some of the matrix elements become zero if the index becomes
negative (the SUSY zero). For the moment we assume that no such zero appears 
in the superpotential.
In general, it is seen that we have  three  massless modes out of the 
six ${\bf \bar 5}$ fields
by solving the above  $3\times 6$ matrix. 
However, since the matrix has hierarchical 
structure, we can easily classify the cases.
\begin{enumerate}
\item  Under the condition that we have no SUSY zeros, 
it is evident that the largest mass is  either $M_{33}$ 
or $M'_{33}$, whose ratio is $M'_{33}/M_{33}=\lambda^r$. 

\item  $0<r$ case: In this case 
 $M_{33}$ is larger than $M'_{33}$, and the pair 
 ($\Psi_3({\bf 10,\bar 5}), \Psi_3({\bf 10,5})$) is heavy. 
Next, compare  $M'_{23}$ 
and  $M_{22}$, whose ratio is
$M'_{23}/M_{22}=\lambda^{r+\psi_3-\psi_2}$.
Thus, there are several cases, depending on $r$ and 
$\psi_3-\psi_2$.

\item $0<r<\psi_2-\psi_3$: In this case, $M'_{23}>M_{22}$, so that 
the pair $\Psi_3({\bf 16,\bar 5})\Psi_2({\bf 10,5})$ becomes heavy,
 and at the
 same time  the pair 
 $\Psi_2({\bf 10,\bar 5})\Psi_1({\bf 10,5})$ obtains a large mass,
because $M'_{12}/M_{12}=\lambda^r<1$. 
$\Psi_1({\bf 10,\bar 5})$, $\Psi_1({\bf 16,\bar 5})$ and 
$\Psi_2({\bf 16,\bar 5})$
are left massless. This case is denoted $(1,1',2)$. 
[In this paper, massless mode whose dominant component is 
$\Psi_i({\bf 16,\bar 5})$ ($\Psi_i({\bf 10,\bar 5})$) is
simply denoted by $i(i')$.]

\item   $(0<)\psi_2-\psi_3<r$: The pair 
$\Psi_2({\bf 10,\bar 5})\Psi_2({\bf 10, 5})$ becomes heavy.
 Further, this case is divided into two cases, according to 
the sign of $r+\psi_3-\psi_1$.

\item $\psi_2-\psi_3<r<\psi_1-\psi_3$: In this case,  
$M'_{13}$ is larger than 
$M_{11}$. Thus  $\Psi_3({\bf 16,\bar 5})\Psi_1({\bf 10, 5})$ 
becomes heavy, and this case also becomes the case $(1,1',2)$. 

\item $(\psi_2-\psi_3<)\psi_1-\psi_3<r$: 
In this  extreme case,  $M'_{13}<M_{11}$, and thus 
 $\Psi_1({\bf 10,\bar 5})\Psi_1({\bf 10, 5})$
becomes heavy. Hence all the 
 $\Psi_i({\bf 10, \bar 5})$ are heavy states and the
 $\Psi_i({\bf 16, \bar 5})$ are massless modes. 
This corresponds to the situation in which the three massless 
${\bf \bar 5}$ 
fields (quarks and leptons) belong to
$\Psi_i({\bf 16, \bar 5})$. This is just the case  usually adopted 
in the $SO(10)$ model. We call this case that of
``parallel family structure.'' We denote  this case  simply
as $(1,2,3)$.

\item $r<0$ case: This case is easily classified just replacing
 the ${\bf 10}$
representation with the  ${\bf 16}$
representation. 
\end{enumerate}
Thus we can classify all the cases as follows: 
\begin{enumerate}
\item $\psi_1-\psi_3<r: (1,2,3)$ type.
\item  $0<r<\psi_1-\psi_3: (1,1',2)$ type.
\item  $\psi_3-\psi_1<r<0: (1,1',2')$ type.
\item  $r<\psi_3-\psi_1: (1',2',3')$ type.
\end{enumerate}
If we use SUSY zero coefficients, various types of massless modes can
be realized.
For example,
if $\psi_1+\psi_3+\phi<0$, SUSY zeros appear, and
the Yukawa terms $\Psi_3\Psi_i\Phi$ $(i=1,2,3)$
are forbidden. Hence the mass matrix $M$ becomes
\begin{equation}
M\rightarrow \bordermatrix{
&\Psi_1({\bf 10, \bar 5})&\Psi_2({\bf 10, \bar5})
&\Psi_3({\bf 10,\bar5})\cr
\Psi_1({\bf 10,5})&\lambda^{2\psi_1}&\lambda^{\psi_1+\psi_2}&0  \cr
\Psi_2({\bf 10,5})& \lambda^{\psi_1+\psi_2}   &  
(\lambda^{2\psi_2})  &0 \cr
\Psi_3({\bf 10,5})  & 0 & 0   & 0  \cr}\lambda^{\phi}v,
\end{equation}
and the massless mode $\Psi_3({\bf 10, \overline 5})$ does not mix 
through  non-diagonal mass
matrix elements with any other ${\bf \bar5}$ field. We call 
such a massless field an ``isolated" field. There are various 
different patterns of massless modes containing the ``isolated" fields.
For example, if the conditions $2\psi_2+\phi\geq 0$, $2\psi_3+c\geq 0$ and
$\lambda^{2\psi_1+\phi} v >\lambda^{\psi_1+\psi_2+c} v'$ are satisfied
in addition
to the above condition $\psi_1+\psi_3+\phi<0$, we have the pattern
$(1,2,3')$, i.e.,
\begin{equation}
\left(\begin{array}{c}
      {\bf \overline 5}_1 \\
      {\bf \overline 5}_2 \\
      {\bf \overline 5}_3
      \end{array} \right)=
\left(\begin{array}{c}
      \Psi_1({\bf 16},{\bf \overline 5})+\cdots \\
      \Psi_2({\bf 16},{\bf \overline 5})+\cdots \\
      \Psi_3({\bf 10},{\bf \overline 5})
      \end{array} \right),
\end{equation}
which has been adopted in Ref. \citen{Bando}.
Note that ${\bf \overline 5}_3$ does not mix with any other states
(an isolated field).

In addition to the mixing of the matter content, 
the massless Higgs doublets itself can in principle be mixed as
\begin{equation}
H({\bf \overline 5})=H({\bf 10},{\bf \overline 5})\cos \theta
                    +H({\bf 16},{\bf \overline 5})\sin \theta,
\label{sin}
\end{equation}
which is also determined by obtaining a whole
mass matrix of the doublet Higgs fields.
Note that the Yukawa couplings 
$\Psi_i({\bf 16},{\bf 10})\Psi_j({\bf 16},{\bf \overline 5})
H({\bf \bar 5})$ 
($\Psi_i({\bf 16},{\bf 10})\Psi_j({\bf 10},{\bf \bar 5})
H({\bf \bar 5}))$
are proportional to $\cos \theta$($\sin \theta$).

\section{Features of the vacua in  the $U(1)_A$ framework}
In this section, we explain how the vacua of the Higgs fields are
determined by the anomalous
$U(1)_A$
quantum numbers.\cite{maekawa}

First, the VEV of a gauge invariant operator with positive
anomalous $U(1)_A$ charge must vanish.
Otherwise, the mechanism of the SUSY zero does not work, since such a
VEV can
compensate for the negative $U(1)_A$ charge of the term.
Generically, such an undesired vacuum is allowed, but as is shown
in the Appendix, in such a vacuum, the Froggatt-Nielsen mechanism\cite{FN}
does not operate.
Therefore we are not
interested in such a vacuum.
Here, we simply assume that we are in the vacuum where the SUSY zero
and Froggatt-Nielsen mechanism operate,
namely any VEV of a gauge invariant operator with positive anomalous
$U(1)_A$ charge vanishes.

Next, we show that
the VEV of a gauge invariant operator $O$ is
determined by its $U(1)_A$ charge $o$ as $\VEV{O}=\lambda^{-o}$
if the $F$-flatness condition determines the VEV.
For simplicity, we examine this relation using singlet fields $Z_i$
with anomalous $U(1)_A$ charge $z_i$.
The general superpotential is written
\begin{eqnarray}
W&=&\sum_i \lambda^{z_i}Z_i+\sum_{i,j}\lambda^{z_i+z_j}Z_iZ_j+\cdots \\
&=&\sum_i \tilde Z_i+\sum_{i,j}\tilde Z_i\tilde Z_j+\cdots,
\end{eqnarray}
where $\tilde Z_i\equiv \lambda^{z_i}Z_i$.
The equations for the $F$-flatness of the $Z_i$ fields require
\begin{equation}
\lambda^{z_i}(1+\sum_j\tilde Z_j+\cdots)=0,
\end{equation}
which generically lead to solutions $\tilde Z_j\sim O(1)$
so that $\VEV{Z_i}\sim \lambda^{-z_i}$, as stated above.
Note that the Froggatt-Nielsen structure of Yukawa couplings,
$\lambda^{\psi_i+\psi_j+h}\Psi_i\Psi_jH$, 
is not changed by the interactions
$W=\lambda^{\psi_i+\psi_j+h+z_k}Z_k\Psi_i\Psi_jH$ with the VEVs
$\VEV{Z_k}\sim \lambda^{-z_k}$. 

If an adjoint field $A$ possesses a non-zero VEV by the $F$-flatness 
condition, this VEV is determined
as $\VEV{A}\sim \lambda^{-a}$, because $A^2$ can be gauge invariant.
Suppose that, in addition to $\Phi$ and $C$, there are
$\bar \Phi({\overline {27}})$ and $\bar C({\overline {27}})$.
Since $\bar \Phi \Phi$ is also gauge invariant, the VEV of the operator is
given by $\VEV{\bar \Phi \Phi}\sim \lambda^{-(\bar \phi+\phi)}$ if it
is determined by the $F$-flatness condition. The $D$-flatness condition of
the $E_6$ gauge group requires
\begin{equation}
|\VEV{\bar \Phi}|=|\VEV{\Phi}|\sim \lambda^{-(\bar \phi+\phi)/2}.
\end{equation}
Note that these VEVs are also determined by the anomalous $U(1)_A$
charges, but they are different from the naive expectation
$\VEV{\Phi}\sim\lambda^{-\phi}$. 
This is because the $D$-flatness condition plays an important
role
in fixing the VEVs.
The VEVs of $C$ and $\bar C$ are also determined by the anomalous $U(1)_A$
charges as
\begin{equation}
|\VEV{C}|=|\VEV{\bar C}|\sim \lambda^{-(c+\bar c)/2}.
\end{equation}
By the above argument, it is found that $v$ and $v'$ are determined
by the anomalous $U(1)_A$ charges. Therefore, the massless modes of 
${\bf \bar 5}$,
which are determined by the twisting mechanism, are also determined by
the anomalous $U(1)_A$ charges.

\section{Quark and lepton masses in $E_6$ unification}
Now let us consider a simple model in which realistic mass matrices of 
quark and lepton are obtained.
Consider the following minimal matter content 
and Higgs  chiral fields. Here, in addition to R-parity,
we introduce $Z_2$ parity, which plays an important role in solving 
the DT splitting problem, as explained in separate 
papers.\cite{maekawa,BKM}
\begin{enumerate}
\item Matter multiplet (odd R-parity):  $\Psi_i({\bf 27},+)$ $i=1,2,3$.
\item Higgs field which breaks $E_6$ into $SO(10)$:
$\Phi({\bf 27},+)$, $\bar \Phi({\bf \overline{27}},+)$,
$\VEV{\Phi}($=$\VEV{\bar \Phi}$).
\item Adjoint Higgs
field $A({\bf 78},-)$, whose $SO(10)$ component $A({\bf 45})$ breaks
$SO(10)$ into
$SU(3)_C\times SU(2)_L\times SU(2)_R\times U(1)_{B-L}$
by the VEV $\VEV{A({\bf 45})}_{B-L}=i\tau_2\times {\rm diag}
(V,V,V,0,0)$. (This Dimopoulos-Wilczek form of the VEV plays an
important role in solving the DT splitting problem.)
\item Higgs field which breaks $SU(2)_R\times U(1)_{B-L}$ into
$U(1)_Y$:
$C({\bf 27},+)$, $\bar C({\bf \overline{27}},+)$
by developing $\VEV{C}($=$\VEV{\bar C}$).
\item Higgs field which contains usual $SU(2)_L$ doublet:
$H({\bf 27},+)$.
\end{enumerate}
In the above, the signature $\pm$ indicates the $Z_2$ parity of the fields. 
Here we have introduced Higgs fields $H$ in addition to the
other Higgs fields $\Phi$, $\bar \Phi$, $C$ and $\bar C$, but it might be
the case that the Higgs doublet can be a part of a component of
the other Higgs fields $\Phi$ and/or $C$.
Even in that case,
the following argument can be applied by taking $h=\phi$ or $h=c$.

In the following, we take the $U(1)_A$ charges of the matter fields
$\Psi_i$ as $\psi_1=3+n$, $\psi_2=2+n$ and $\psi_3=n$, 
which have been determined
in previous papers to be consistent with the up-type quark masses and
mixings.
Then the top Yukawa coupling of order 1 determines the anomalous
$U(1)_A$
charge of the Higgs field $H$ as $h=-2n$.

Also, in this paper, we assume that the mixing angle $\sin \theta$
[defined in Eq.(\ref{sin})]
is zero, i.e., the down-type Higgs is purely
$H({\bf 10},{\bf \overline 5})$. This assumption makes the
following analysis much simpler. 
Of course, once we determine the model that realizes the DT splitting, 
the Higgs mixing 
angle is also determined by the anomalous $U(1)_A$ charges.
We shall discuss this point in a separate paper.\cite{BKM}
Actually, we find various DT splitting models that give
$\sin \theta=0$.

Now the Yukawa couplings are obtained by Froggatt-Nielsen 
mechanism\cite{FN} as
\begin{equation}
W=\lambda^{\psi_i+\psi_j+h}\Psi_i\Psi_jH,
\label{yukawa}
\end{equation}
where the mass matrix of the up quark sector is uniquely determined,
since we have already fixed the $U(1)$ charges of the fields
$\Psi_i({\bf 27})$:
\begin{equation}
M_u=
\bordermatrix{
&\Psi_1({\bf 16,10})&\Psi_2({\bf 16,10})&\Psi_3({\bf 16,10}) \cr
\Psi_1({\bf 16,10})&\lambda^6 & \lambda^5 & \lambda^3 \cr
\Psi_2({\bf 16,10})&\lambda^5 & \lambda^4 & \lambda^2 \cr
\Psi_3({\bf 16,10})&\lambda^3 & \lambda^2 & 1 \cr}
\VEV{H({\bf 10,5})}.
\end{equation}
The twisting mechanism discussed in
the section 2 causes the down quark mass matrix
to differ from that of the  up quark.

We examine the massless modes of ${\bf \overline 5}$ in the
following under the assumption 
 $\sin \theta=0$.
Note that in such a situation, component fields
$\Psi_i({\bf 10,\overline 5})$,
have no Yukawa couplings, because the Yukawa terms
$\Psi_i({\bf 10,\overline 5})\Psi({\bf 16,10})H({\bf 10,\overline 5})$
are forbidden by $SO(10)$ gauge symmetry.
This excludes the cases that include isolated
$\Psi_i({\bf 10,\overline 5})$ fields. Moreover, the cases that
include isolated $\Psi_i({\bf 16,\overline 5})$ fields should be discarded,
since we cannot obtain large neutrino mixing angles in such cases.
With the charge assignment $\psi_1=n+3$, $\psi_2=n+2$ and $\psi_3=n$,
the classification discussed in section 2 
applies even to cases in which SUSY zeros appear, provided that there
are no isolated fields.
Among the options $(1,2,3)$, $(1,1',2)$, $(1,1',2')$
and $(1',2',3')$, only the option $(1,1',2)$ gives realistic quark and
lepton mass matrices.

Let us examine an case of $(1,1',2)$, i.e., $0<r<3$,
with the constraints
that forbid the existence of an isolated state,
$0\leq \psi_1+\psi_3+c$ and
$0\leq \psi_1+\psi_3+\phi$.
Here, the parameter $r$, which has been defined by
$\lambda^r=\lambda^cv'/(\lambda^\phi v)$, is given by
\begin{equation}
r=\frac{1}{2}[c-\bar c-(\phi-\bar \phi)],
\label{r}
\end{equation}
because the VEVs
$v$ and $v'$ are fixed by the anomalous $U(1)_A$ charges.
Note that even if we take the anomalous $U(1)_A$ charges as integers,
$r$ can be a half-integer. This fact plays an important role
in realizing the bi-large neutrino mixing angles, as we see in the
next section.
With this case, we investigate which type of mixing pattern of
$\overline 5$ fields can
reproduce the bi-large neutrino mixing.
In order to see this, let us consider the cases 
$({\bf \overline 5}_1,{\bf \overline 5}_2,{\bf \overline 5}_3)
=(1,1',2)$ and 
$({\bf \overline 5}_1,{\bf \overline 5}_2,{\bf \overline 5}_3)
=(1,2,1')$
as phenomenologically viable
patterns of the massless three fields 
$({\bf \overline 5}_1,{\bf \overline 5}_2,{\bf \overline 5}_3)$.
Note that the correct expression of the
massless states ${\bf \bar 5}_i$ at low energy are obtained
as mixed states of $\Psi_j$ by solving
the mass matrix of Eq.(\ref{full}).
It should be remarked that
$\Psi_1({\bf 10},{\bf \overline 5})$ itself
does not have a Yukawa coupling, and therefore the field $1'$ really
can have a Yukawa coupling only through the mixing with
$\Psi_i({\bf 16},{\bf \overline 5})$.
In order to obtain the exact mass matrix for down quarks
as well as leptons, we should take account of the
mixing effects from the non-dominant states.
We first fix the three bases of the massless modes
$({\bf \overline 5}_1,{\bf \overline 5}_2,{\bf \overline 5}_3)$
to
$(\Psi_1({\bf 16},{\bf \overline 5}),
\Psi_1({\bf 10},{\bf \overline 5}),
\Psi_2({\bf 16},{\bf \overline 5}))$.
On this basis, we can estimate the order of mixing parameters with
the heavy states
$\Psi_3({\bf 16},{\bf \overline 5})$,
$\Psi_2({\bf 10},{\bf \overline 5})$
and $\Psi_3({\bf 10},{\bf \overline 5})$ as
\begin{eqnarray}
{\bf \overline 5}_1 &=& \Psi_1({\bf 16},{\bf \overline 5})
+\lambda^{\psi_1-\psi_3}\Psi_3({\bf 16},{\bf \overline 5})
+\lambda^{\psi_1-\psi_2+r}\Psi_2({\bf 10},{\bf \overline 5})
+\lambda^{\psi_1-\psi_3+r}\Psi_3({\bf 10},{\bf \overline 5}), 
\label{51} \\
{\bf \overline 5}_2 &=& \Psi_1({\bf 10},{\bf \overline 5})
+\lambda^{\psi_1-\psi_3-r}\Psi_3({\bf 16},{\bf \overline 5})
+\lambda^{\psi_1-\psi_2}\Psi_2({\bf 10},{\bf \overline 5})
+\lambda^{\psi_1-\psi_3}\Psi_3({\bf 10},{\bf \overline 5}), 
\label{52} \\
{\bf \overline 5}_3 &=& \Psi_2({\bf 16},{\bf \overline 5})
+\lambda^{\psi_2-\psi_3}\Psi_3({\bf 16},{\bf \overline 5})
+\lambda^{r}\Psi_2({\bf 10},{\bf \overline 5})
+\lambda^{\psi_2-\psi_3+r}\Psi_3({\bf 10},{\bf \overline 5}),
\label{53}
\end{eqnarray}
where the first terms on the right-hand sides are the main components of
these massless modes, and the other terms are mixing terms with
heavy states, $\Psi_3({\bf 16},{\bf \overline 5})$,
$\Psi_2({\bf 10},{\bf \overline 5})$ and
$\Psi_3({\bf 10},{\bf \overline 5})$.
The order of these mixing parameters can be estimated by
the ratios of the relevant mass matrix elements.
For example, the ratio of the mass matrix element
$M'_{k1}=\lambda^{\psi_1+\psi_k+c}v'$ to
$M'_{k3}=\lambda^{\psi_3+\psi_k+c}v'$ becomes
$M'_{k1}/M'_{k3}=\lambda^{\psi_1-\psi_3}$, which
appears in the coefficient of the second term of Eq. (\ref{51}).
Note that this ratio is independent of the parameter $\psi_k$.
Similarly, the ratio of $M'_{k1}=\lambda^{\psi_1+\psi_k+c}v'$
to $M_{k3}=\lambda^{\psi_3+\psi_k+\phi}v$ becomes
$M'_{k1}/M_{k3}=\lambda^{\psi_1-\psi_3+r}$,
which appears in the coefficient of the third term in Eq. (\ref{51}).

The mass matrices of the down-type quark and charged lepton can be obtained
from the above mixing pattern by introducing the RGE factor 
$\eta^{-1}\sim$ 2--3. We then have
\begin{equation}
M_D=M^T_E\eta^{-1}=\bordermatrix{
 & {\bf \overline 5}_1 &{\bf \overline 5}_2 &{\bf \overline 5}_3 \cr
\Psi_1({\bf 16,10})  &  \lambda^6     &  \lambda^{6-r}  & \lambda^5  \cr
\Psi_2({\bf 16,10})  &  \lambda^5     &  \lambda^{5-r}  & \lambda^4 \cr
\Psi_3({\bf 16,10})  &  \lambda^3     &  \lambda^{3-r}  & \lambda^2
 \cr}\VEV{H({\bf 10,\overline 5})},
\end{equation}
which corresponds to the case $(1,1',2)$, for which
$3-r>2\,\rightarrow \, 1>r$.\footnote{
If $1<r$, the second family should be
exchanged with the third family (the case $(1,2,1')$).}
Note that in Eq. (\ref{52}), the main mode of ${\bf \overline 5}_2$ is
$\Psi_1({\bf 10,\overline 5})$, which has no Yukawa coupling
to $H({\bf 10,\overline 5})$. Therefore the contribution from the
mixing term
$\lambda^{\psi_1-\psi_3-r}\Psi_3({\bf 16},{\bf \overline 5})$
determines the order of the Yukawa couplings.
On the other hand, the main modes of ${\bf \overline 5}_1$ and
${\bf \overline 5}_3$ determine the order of the Yukawa couplings,
while the contribution of the mixing terms is of the same order.

Now that we have the mass matrices for up and down quarks, we can
 estimate the CKM matrix\footnote{
Strictly speaking, if the Yukawa coupling originates only from the 
interaction (\ref{yukawa}), the mixing involving the first generation
becomes too small, 
due to a cancellation. In order to obtain the expected value of
the CKM matrix as in Eq. (\ref{CKM}), non-renormalizable terms, for
example $\Psi_i\Psi_jH\bar CC$, must be taken into account.
}
as
\begin{equation}
U_{\rm CKM}=
\left(
\begin{array}{ccc}
1 & \lambda &  \lambda^3 \\
\lambda & 1 & \lambda^2 \\
\lambda^3 & \lambda^2 & 1
\end{array}
\right),
\label{CKM}
\end{equation}
which is consistent with the experimental value if we take
$\lambda\sim 0.2$.
Since the ratio of the Yukawa couplings of top and bottom quarks is
$\lambda^2$,
a small value of 
$\tan \beta\equiv \VEV{H({\bf 10,5})}/\VEV{H({\bf 10,\overline 5})}
\sim m_t/m_b\cdot \lambda^2$ is
predicted
by these mass matrices.
The Yukawa matrix for the charged lepton sector is the same as the transpose
of $M_d$ at this stage, except for an overall factor $\eta$ induced by the
renormalization group effect.

\section{Bi-large neutrino mixing in $E_6$ unification}
Now we treat the neutrino masses and mixing. In order to do this,
we must estimate 
the mixings in the neutrino mass matrix, since the Maki-Nakagawa-Sakata (MNS)
matrix\cite{MNS} is given by
\begin{equation}
  U_{\rm MNS}=U_lU_\nu^\dagger\ ,
\end{equation}
with the unitary matrices $ U_l$ and $U_\nu$  that make the matrices
$U_E(M_E^\dagger M_E)U_E^\dagger$
and $U_\nu^*M_\nu U_\nu^\dagger$ diagonal. The matrix $M_\nu$ is the
Majorana mass matrix of the light (almost) left-handed neutrinos,
which is obtained from the Dirac masses and right-handed Majorana
masses.
First, the Dirac neutrino mass matrix is given by the $3\times 6$ matrix
\begin{equation}
\bordermatrix{
    &\Psi_1({\bf 1,1})&\Psi_2({\bf 1,1})&\Psi_3({\bf1,1})
    &\Psi_1({\bf 16,1})&\Psi_2({\bf 16,1}) &\Psi_3({\bf16,1})   \cr
{\bf \overline 5}_1 &  \lambda^{r+6}  &  \lambda^{r+5}   & \lambda^{r+3}  
&  \lambda^6 &  \lambda^5     & \lambda^3  \cr
{\bf \overline 5}_2  &  \lambda^6   &  \lambda^5  &\lambda^3 &  
\lambda^{6-r} &  \lambda^{5-r}&\lambda^{3-r}\cr
{\bf \overline 5}_3  & \lambda^{r+5}   &  \lambda^{r+4}  & \lambda^{r+2}
&  \lambda^5   &\lambda^4  &\lambda^2
 \cr}\VEV{H({\bf 10,5})}\eta,
\end{equation}
or we simply express it as
\begin{equation}
M_{N}=
\left(
\begin{array}{cc}
 \lambda^{r+2} & \lambda^2
\end{array}
\right)\otimes
\left(
\begin{array}{ccc}
\lambda^4 & \lambda^3 & \lambda \\
\lambda^{4-r} & \lambda^{3-r} & \lambda^{1-r} \\
\lambda^3   & \lambda^2         & 1
\end{array}
\right)\VEV{H({\bf 10,5})}\eta.
\end{equation}
The right-handed Majorana masses come from the
interaction
\begin{equation}
\lambda^{\psi_i+\psi_j+2\bar \phi}\Psi_i\Psi_j\bar \Phi\bar \Phi
+\lambda^{\psi_i+\psi_j+\bar c+\bar \phi}\Psi_i\Psi_j\bar \Phi\bar C
+\lambda^{\psi_i+\psi_j+2\bar c}\Psi_i\Psi_j\bar C\bar C.
\end{equation}
Then, the $6\times 6$ matrix for
$\Psi_i({\bf 1,1}),i=1,2,3$, and $\Psi_k({\bf 16,1}),k=1,2,3$, 
the right-handed neutrinos, is expressed as 
\begin{eqnarray}
M_R&=
&\lambda^{\psi_i+\psi_j+2\bar \phi}\Psi_i({\bf 1,1})\Psi_j({\bf 1,1})
\VEV{\bar \Phi}^2+\lambda^{\psi_i+\psi_k
+\bar c+\bar \phi}\Psi_i({\bf 1,1})\Psi_k({\bf 16,1})
\VEV{\bar \Phi}\VEV{\bar C} \nonumber \\
&&+\lambda^{\psi_k+\psi_m+2\bar c}\Psi_k({\bf 16,1})\Psi_m({\bf 16,1})
\VEV{\bar C}^2 \\
&=&\lambda^{2n}
\left(
\begin{array}{cc}
\lambda^{\bar \phi-\phi} & \lambda^{(\bar \phi-\phi+\bar c-c)/2} \\
\lambda^{(\bar \phi-\phi+\bar c-c)/2} & \lambda^{\bar c-c}
\end{array}
\right)
\otimes
\left(
\begin{array}{ccc}
\lambda^6 & \lambda^5 & \lambda^3 \\
\lambda^5 & \lambda^4 & \lambda^2 \\
\lambda^3   & \lambda^2         & 1
\end{array}
\right),
\end{eqnarray}
from which the neutrino mass matrix is found using the seesaw mechanism
\cite{seesaw} to be
\begin{equation}
M_\nu=M_NM_R^{-1}M_N^T=\lambda^{4-2n+c-\bar c}\left(
\begin{array}{ccc}
\lambda^2 & \lambda^{2-r} & \lambda \\
\lambda^{2-r} & \lambda^{2-2r} & \lambda^{1-r} \\
\lambda   & \lambda^{1-r}         & 1
\end{array}
\right)\VEV{H({\bf 10,5})}^2\eta^2,
\end{equation}
where we have used the relation (\ref{r}).

Combining the charged lepton sector from the previous section 
and neutrino sector from above, we finally
obtain the Maki-Nakagawa-Sakata matrix:\footnote{
In the case  $r>1$ $(1,2,1')$, we obtain
\begin{equation}
U_{\rm MNS}=
\left(
\begin{array}{ccc}
1 & \lambda &  \lambda^{r} \\
\lambda & 1 & \lambda^{r-1} \\
\lambda^{r} & \lambda^{r-1} & 1
\end{array}
\right). \nonumber
\end{equation}
}
\begin{equation}
U_{\rm MNS}=
\left(
\begin{array}{ccc}
1 & \lambda^r &  \lambda \\
\lambda^{r} & 1 & \lambda^{1-r} \\
\lambda & \lambda^{1-r} & 1
\end{array}
\right).
\end{equation}

Recent experiments on atmospheric neutrinos have suggested
a very large mixing angle
between
second and third generations, and thus
$r=1/2,1$ may be  realistic [for the case of $(1,1',2)$, i.e.,  
$r \leq 1$].\footnote{
In the case of $(1,2,1')$, the parameter value $r=3/2$ may yield a
prediction consistent with the large mixing indicated by atmospheric
neutrino experiments.}
It turns out that  $r=1/2$ actually leads to
bi-large neutrino mixing angles, which are examined within the
$SO(10)$ model in Ref. \citen{maekawa}.\footnote{
When $r=1$, the fermion mass matrices become of the ``lopsided"
type. This would seem to give a small mixing angle solution for the solar 
neutrino problem. However, recently it has been pointed 
out that taking account $O(1)$ coefficients, lopsided-type 
mass matrices can give even large
mixing angle solutions for solar neutrino problem. }
Indeed if we take $r=1/2$, namely,
\begin{equation}
c-\bar c=\phi-\bar \phi+1,
\label{rr}
\end{equation}
the MNS matrix is given by
\begin{equation}
U_{\rm MNS}=
\left(
\begin{array}{ccc}
1 & \lambda^{1/2} &  \lambda \\
\lambda^{1/2} & 1 & \lambda^{1/2} \\
\lambda & \lambda^{1/2} & 1
\end{array}
\right),
\end{equation}
which gives bi-large mixing angles for the neutrino sector,
since $\lambda^{1/2}\sim 0.5$. At the same time it predicts
$V_{e3}\sim \lambda$. It will be interesting to see if
future experiments find evidence just below the  CHOOZ  upper
limit $V_{e3}\leq 0.15$.
\cite{CHOOZ}
For the neutrino masses, the model predicts
$m_{\nu_\mu}/m_{\nu_\tau}\sim \lambda$,
which is consistent with
the experimental data:
$1.6\times 10^{-3} {\rm eV}^2\leq \Delta m_{\rm atm}^2\leq 4
\times 10^{-3}
{\rm eV}^2$
and $2\times 10^{-5} {\rm eV}^2\leq \Delta m_{\rm solar}^2\leq 1
\times 10^{-4}
{\rm eV}^2$, which is the allowed region for the most probable
MSW solution for the solar neutrino (LMA).
\cite{SK}

If we enforce the condition
\begin{equation}
\phi-\bar\phi=2n-10-l,
\end{equation}
the neutrino mass matrix is obtained as
\begin{equation}
M_\nu=\lambda^{-(5+l)}\left(
\begin{array}{ccc}
\lambda^2 & \lambda^{2-r} & \lambda \\
\lambda^{2-r} & \lambda^{2-2r} & \lambda^{1-r} \\
\lambda   & \lambda^{1-r}         & 1
\end{array}
\right)\VEV{H({\bf 10,5})}^2\eta^2,
\end{equation}
where we have used the relation (\ref{rr}). From the above equation, 
we obtain
\begin{equation}
\lambda^l=\lambda^{-5}\frac{\VEV{H({\bf 10,5})}^2\eta^2}{m_{\nu_\tau}\Mp}.
\end{equation}
We are supposing that the cutoff scale $\Mp$ is in the range
$10^{16}{\rm GeV}<\Mp<10^{20}{\rm GeV}$, which allows
$-2\leq l \leq 2$. 
If we choose $l=0$,
the neutrino masses are given  by
$m_{\nu_\tau}\sim \lambda^{-5}\VEV{H({\bf 10,5})}^2\eta^2/\Mp\sim
m_{\nu_\mu}/\lambda
\sim m_{\nu_e}/\lambda^2$. If we take $\eta\VEV{H({\bf 10,5})}=100$ GeV,
$\Mp\sim 10^{18}$ GeV and $\lambda=0.2$, then we get
$m_{\nu_\tau}\sim 3\times 10^{-2}$ eV, $m_{\nu_\mu}\sim 6\times
10^{-3}$ eV
and $m_{\nu_e}\sim 1\times 10^{-3}$ eV. From such a rough estimation, 
we can obtain values that are nearly consistent with 
the experimental data for atmospheric neutrinos and we can also obtain
a large mixing angle (LMA) MSW solution for the solar neutrino 
problem.\cite{MSW} This LMA solution for the solar neutrino problem
gives the best to the present experimental data.\cite{Valle}

Finally, we would like to make a comment on an interesting feature
of this scenario, which is also seen in the $SO(10)$ model.\cite{maekawa}
 In addition to Eq.(\ref{yukawa}), the interactions
\begin{equation}
\lambda^{\psi_i+\psi_j+2a+h}\Psi_iA^2\Psi_jH
\end{equation}
also contribute to the Yukawa couplings after $A$ develops a non-vanishing
VEV. Here, only $A^2$ appears
because of its odd $Z_2$ parity.
Since $\VEV{A}$ is proportional to the generator of $B-L$,
the contribution to the lepton Yukawa coupling is nine times larger
than that to the quark Yukawa couplings.
If we set  $a=-2$,
the additional matrices are
\begin{eqnarray}
\frac{\Delta M_u}{\VEV{H({\bf 10,5})}}&=&
\frac{V^2}{4}\left(
\begin{array}{ccc}
\lambda^2 & \lambda & 0 \\
\lambda & 1 & 0 \\
0 & 0 & 0
\end{array}
\right)
,\quad
\frac{\Delta M_d}{\VEV{H({\bf 10,\overline 5})}}=
\frac{V^2}{4}\left(
\begin{array}{ccc}
\lambda^2 & 0 & \lambda \\
\lambda & 0 & 1 \\
0 & 0 & 0
\end{array}
\right)
,
\\
\frac{\Delta M_e}{\VEV{H({\bf 10,\overline 5})}}&=&
\frac{9V^2}{4}\left(
\begin{array}{ccc}
\lambda^2 & \lambda & 0 \\
0 & 0 & 0 \\
\lambda & 1 & 0
\end{array}
\right).
\end{eqnarray}
Note that the additional terms contribute mainly to the lepton
sector.
It is interesting that this modification essentially changes the
mass eigenvalues of
only the first and second generation. Hence it is natural to expect that
a realistic mass pattern can be obtained by this modification:
It changes  the unrealistic
prediction $m_\mu=m_s$ at the GUT scale
while preserving the beautiful prediction $m_b=m_\tau$
at the GUT scale (the GUT relation).\footnote{
Strictly speaking that are forbidden by the SUSY zero mechanism
 are generically induced by integrating out
 heavy fields that are introduced to solve the  DT splitting
 problem. These terms may give a small correction to the GUT
 relation $m_b=m_\tau$.}
This enhancement factor of $2-3$ of $m_\mu$ can be enough to
improve the unwanted situation of the lepton quark relation
in the second family.

Remarkably enough, this charge assignment of $A$
 determines the  scale of $\VEV{A}$ as $\sim \lambda^2$.
This strong correlation of the unification scale, which
 is a bit
smaller than the Planck scale, and the improvement of
the undesired GUT prediction $m_\mu=m_s$ is indeed a
consequence of $U(1)_A$.
It is also interesting that the SUSY zero plays an essential role again.
When $z, \bar z \geq -4$, the terms
$\lambda^{\psi_i+\psi_j+a+z+h}Z\Psi_iA\Psi_jH+
\lambda^{\psi_i+\psi_j+2z+h}Z^2\Psi_i\Psi_jH$
also contribute to the fermion mass matrices, though only to the first
generation.

\section{SUSY breaking and FCNC}
Finally, we discuss SUSY breaking. Since we should choose the
anomalous $U(1)_A$ charges dependent on the flavour to produce the
hierarchy of Yukawa couplings, generically
non-degenerate scalar fermion masses are induced through the anomalous
$U(1)_A$ $D$-term.\footnote{
The large SUSY breaking scale can make it possible to avoid the flavour 
changing
neutral current (FCNC) problem,\cite{DG,CKN} but in our scenario
this is not the case, because the anomalous $U(1)_A$ charge of the Higgs
$H$ is
inevitably negative, which forbids the Higgs mass term at the tree level.
}
Various experiments on FCNC processes give
strong constraints on
the off-diagonal terms $\Delta$ in the sfermion mass matrices
due to the fact that the flavour-changing terms appear only in the 
off-diagonal parts of the sfermion propagators, as seen in 
Ref. \citen{masiero}.
The sfermion propagators can be expanded in terms of 
$\delta=\Delta/\tilde m^2$, where $\tilde m$ is the average sfermion mass.
As long as $\Delta$ is sufficiently smaller than $\tilde m^2$, 
it is enough to take the first term of this expansion, and then, 
the experimental
information concerning FCNC and CP violating phenomena is translated into
the upper bounds on these $(\delta_{ij}^F)_{XY}$, where
$F=U,D,N,E$, the chirality index is $X,Y=L,R$, 
and the generation index is $i,j=1,2,3$.
For example, 
the experimental value of $K^0-\bar K^0$ mixing gives
\begin{eqnarray}
&&\sqrt{|{\rm Re} (\delta_{12}^D)_{LL}(\delta_{12}^D)_{RR}|}
\leq 2.8\times 10^{-3}
\left( \frac{\tilde m_q ({\rm GeV})}{500}\right), \label{LR}\\
 &&|{\rm Re} (\delta_{12}^D)_{LL}|,|{\rm Re} (\delta_{12}^D)_{RR}|
 \leq 4.0\times 10^{-2}
\left( \frac{\tilde m_q ({\rm GeV})}{500}\right), \label{LL}
\end{eqnarray}
with $\tilde m_q$ the average value of the squark masses.\footnote{
The CP violation parameter $\epsilon_K$ gives 
constraints on the imaginary part of $(\delta_{12}^D)_{XY}$ 
that are approximately one order more severe than those it places on
the real part.
Here we concentrate only on the constraints from the real
part of $K^0\bar K^0$ mixing, 
since under the 
other experimental constraints on the CP phase originating from 
SUSY breaking sector, which are mainly given by the electric dipole moment,
we may expect that the CP phases are small enough to satisfy the 
constraints from the imaginary part of the $K^0\bar K^0$ mixing.
}
The $\mu\rightarrow e \gamma$ process gives
\begin{equation}
\ |(\delta_{12}^E)_{LL}|,|(\delta_{12}^E)_{RR}|\leq 3.8\times 10^{-3}
\left( \frac{\tilde m_l ({\rm GeV})}{100}\right)^2,\label{LLl}
\label{LFV}
\end{equation}
where $\tilde m_l$ is the average mass of the scalar leptons.
In the usual anomalous $U(1)_A$ scenario, $\Delta$ can be
estimated as
\begin{equation}
(\Delta_{ij}^F)_{XX}\sim
\lambda^{|f_i-f_j|}(|f_i-f_j|)\VEV{D_A},
\end{equation}
since the mass difference is given by $(f_i-f_j)\VEV{D_A}$,
where $f_i$ is the anomalous $U(1)_A$ charge of $F_i$.
Here, the reason for the appearance of the coefficient $\lambda^{|f_i-f_j|}$ 
is that the unitary diagonalizing
matrices are given by
\begin{equation}
\left(
\begin{array}{cc}
1 & \lambda^{|f_i-f_j|} \\
-\lambda^{|f_i-f_j|} & 1
\end{array}
\right).
\end{equation}
In our scenario, the anomalous $U(1)_A$ charge of ${\bf \bar 5}_1$
is the same as that of ${\bf \bar 5}_2$; i.e., the sfermion masses of
${\bf \bar 5}_1$ and 
${\bf \bar 5}_2$ are almost equal. 
This  weakens the constraints from these FCNC processes. 
This is because the constraints from $K^0-\bar K^0$ 
mixing and CP violation on the product 
$(\delta_{12})_{LL}\times (\delta_{12})_{RR}$ are much stronger
than those on $(\delta_{12})_{LL}^2$ or $(\delta_{12})_{RR}^2$,
as shown in Eqs. (\ref{LR}) and (\ref{LL}).
Therefore, suppression of $(\Delta_{12}^D)_{RR}$ makes the constraints
much weaker.
Because the constraints from the $K^0\bar K^0$ mixing (and the 
CP violation) become weaker, as discussed above, we have a larger region
in the  
paramter space where lepton flavour
violating processes like $\mu\rightarrow e \gamma$ are appreciable.
Actually, if the ratio of the VEV of $D_A$ to the gaugino mass squared 
at the GUT scale is given by
\begin{equation}
R\equiv \frac{\VEV{D_A}}{M_{1/2}^2},
\end{equation}
then the scalar fermion mass square at low energy scales is estimated as
\begin{equation}
\tilde m_{F_i}^2\sim f_i R M_{1/2}^2+\eta_FM_{1/2}^2,
\end{equation}
where $\eta_F$ is a renormalization group factor.
Therefore, in our scenario, Eq. (\ref{LL}) for 
$(\delta_{12}^D)_{LL}$
becomes
\begin{eqnarray}
(\delta_{12}^D)_{LL}&\sim &
\lambda\frac{(\psi_1-\psi_2)R M_{1/2}^2}
{(\eta_{D_L}+\frac{\psi_1+\psi_2}{2}R)M_{1/2}^2}
=\lambda\frac{(\psi_1-\psi_2)R}{(\eta_{D_L}+\frac{\psi_1+\psi_2}{2}R)} \\
&\leq &4.0\times 10^{-2}
\left(\frac{(\eta_{D_L}+\frac{\psi_1+\psi_2}{2}R)^{1/2}M_{1/2}({\rm GeV})}
           {500}\right),
\end{eqnarray}
which can be rewritten
\begin{equation}
M_{1/2}\geq 
1.25\times 10^4 \lambda\frac{(\psi_1-\psi_2)R}
{(\eta_{D_L}+\frac{\psi_1+\psi_2}{2}R)^{3/2}}
({\rm GeV}).
\end{equation}
Though the main contribution to $(\delta_{12}^D)_{RR}$ vanishes, through
the mixing in Eqs. (\ref{51}) and (\ref{52}), $(\delta_{12}^D)_{RR}$ is
estimated as\footnote{
We thank S. Yamashita for pointing out the contribution from the
normalization factor of the main mode.}
\begin{equation}
(\delta_{12}^D)_{RR}\sim \lambda^{\frac{1}{2}}\frac{\lambda^2(\psi_1-\psi_2)R}
                                                  {\eta_{D_R}+\psi_1R},
\label{RR}
\end{equation}
where the mixing $\lambda^{\frac{1}{2}}$ is different from the naively
expected value $1=\lambda^{\psi_1-\psi_1}$.
From Eq. (\ref{LR}) for $\sqrt{(\delta_{12}^D)_{LL}(\delta_{12}^D)_{RR}}$,
the constraint on the gaugino mass $M_{1/2}$ is given by
\begin{equation}
M_{1/2}\geq 1.8\times 10^5\frac{\lambda^{1.75} R(\psi_1-\psi_2)}
                               {(\eta_D +\psi_1 R)^{1.5}}.
\end{equation}
On the other hand, Eq. (\ref{LFV}) for $(\delta_{12}^E)_{RR}$ leads to
\begin{equation}
M_{1/2}\geq 1.6\times 10^3
\frac{(\lambda (\psi_1-\psi_2)R)^{1/2}}
{\eta_{E_R}+\frac{\psi_1+\psi_2}{2}R}
({\rm GeV}).
\end{equation}
Taking the reasonable values $\psi_1=5$, $\psi_2=4$, 
$\eta_{D_L}\sim \eta_{D_R}\sim 6$ and 
$\eta_{E_R}\sim 0.15$, 
the lower limits of the gaugino mass are roughly given as in
Table I.

\vspace{3mm}
\begin{center}
Table I. Lower bound of gaugino mass $M_{1/2}$ at the GUT scale (in GeV).
\begin{tabular}{|c|c|c|c|c|c|} 
\hline
$R$                              & 0.1  & 0.3 & 0.5 & 1 & 2 \\ \hline
$(\delta_{12}^D)_{LL}$   & 15 & 38  & 53  & 73 & 86 \\ \hline
$\sqrt{(\delta_{12}^D)_{LL}(\delta_{12}^D)_{RR}}$   
                         & 60 & 150 & 210 & 280 & 350 \\ \hline
$|(\delta_{12}^E)_{RR}| $          & 370 & 260 & 210 & 150 & 110\\ \hline
\end{tabular}

\vspace{5mm}
\end{center}
Note that when $R<0.5$, the $\mu\rightarrow e\gamma$
process gives the severest constraint in these FCNC 
processes.\cite{kurosawa}
We conclude that the lepton flavour violating processes\cite{kurosawa,LFV} 
might be seen in the near future.

The reason for the suppression of $(\Delta_{12}^D)_{RR}$
is that the anomalous $U(1)_A$ charge of ${\bf \overline 5}_2$ becomes the
same as that of ${\bf \overline 5}_1$, because the fields ${\bf \overline
5}_1$
and ${\bf \overline 5}_2$ originate from a single field, $\Psi_1$.
This is a non-trivial situation. The massless mode of the second 
generation
${\bf \overline 5}_2=\Psi_1({\bf 10},{\bf \bar 5})
+\lambda^{5/2}\Psi_3({\bf 16},{\bf \bar 5})$
has Yukawa couplings through the second term
$\lambda^{5/2}\Psi_3({\bf 16},{\bf \bar 5})$. However, for the SUSY breaking
term,
which is proportional to the anomalous $U(1)_A$ charge, the contribution
from the first term dominates that from the second term. 
This results in 
degenerate SUSY breaking terms between the first and the second generation.
It is obvious that the twisting mechanism in $E_6$ unification plays an
essential role in realizing this non-trivial structure.
Note that such a structure is realized only when 
$({\bf \bar 5}_1, {\bf \bar 5}_2)=(1,1')$,\footnote{
The case $(1,1',2')$ cannot realistically yield the large mixing
angle indicated by atmospheric neutrino experiments.
}
in which
bi-large neutrino mixing angles are also realized.
It is suggestive that the requirement to reproduce the bi-large mixing
angles in the neutrino sector leads to this non-trivial structure, which
suppresses the FCNC processes.\footnote{
We should comment on the $D$-term contribution to the scalar fermion
masses. Generically, such a $D$-term has non-vanishing VEV
\cite{kawamura} when the rank of
the gauge group is reduced by the symmetry breaking and SUSY breaking terms
are non-universal. In our scenario,
when the $E_6$ gauge group is broken to the $SO(10)$ gauge group, the $D$-term
contribution gives different values to the sfermion masses of {\bf 16} 
and {\bf 10} of $SO(10)$. This destroys the natural suppression of FCNC
in the $E_6$ unification. However, if SUSY breaking parameters become
universal for some reason, the VEV of $D$ can become negligible. 
Actually, the condition $m_\phi^2=m_{\bar \phi}^2$ causes the VEV of
the $D$ to be greatly suppressed.
Therefore, in principle, we can control the $D$-term contribution, though
it is dependent on the SUSY breaking mechanism. 
}
In this way, such a non-trivial structure is automatically
obtained in the $E_6$ model, which is much different from 
the situation for the $SO(10)$ model, in which the condition
can be satisfied only by hand.

\section{Discussions and summary}
In this paper, we examined an $E_6$ unified model in which bi-large
neutrino
mixing angles are realized. 
A noteworthy fact of such GUT model with the anomalous $U(1)_A$
framework, is that once we fix the charges of all the fields of the
model, all the hierarchical scales, the symmetry breaking scales 
at high energy and also the hierarchical structure of the Yukawa couplings,
are determined 
without any ambiguity. The only exceptions are the SUSY breaking scale and
 the  electroweak breaking 
scale, which we here adjusted from the experimental W masses. 
Even if the SUSY breaking scale is introduced by hand,
we have to explain why the SUSY Higgs mass parameter $\mu$ is around
the SUSY breaking scale (the $\mu$ problem). 
One possible solution of the $\mu$ problem has recently been 
examined in Ref. \citen{maekawa2}.
Here we summarize the essence of the finding there.
The SUSY Higgs mass, which is forbidden by the SUSY zero 
mechanism,
can be induced when SUSY is broken. Thus the parameter $\mu$ must be 
proportional to a
SUSY breaking parameter, 
and its coefficient is determined by anomalous $U(1)_A$
charges.
Let us introduce the GUT gauge singlets $S$, with positive $s$, and
$Z$, with negative $z$, with the mass term $\lambda^{s+z}SZ$ in the
superpotential. Since $S$ has positive charge, it has vanishing
VEV in SUSY vacua (see the Appendix).
When SUSY is broken, generically a tadpole
term $\lambda^sAS$ ($A$ is a SUSY breaking parameter) is induced 
in the SUSY breaking potential $V_{SB}$.
As a result, the $S$ field develops a non-vanishing VEV as
\begin{equation}
\VEV{S}=\lambda^{-s-2z}A.
\end{equation}
Using this VEV shift, we generically find
the $\mu$ term to be proportional to the SUSY breaking paramter $A$.
In our $E_6$ scenario, introducing the superpotential
\begin{equation}
W=\lambda^{s+\phi+2h}S\Phi H^2,
\end{equation}
the SUSY Higgs mass $\mu$ is obtained as
\begin{equation}
\mu\sim \lambda^{2(h-z)+\frac{1}{2}(\phi-\bar \phi)}A.
\end{equation}
Therefore, if 
\begin{equation}
-1\leq 2(h-z)+\frac{1}{2}(\phi-\bar \phi)\leq 1
\label{mu},
\end{equation}
the $\mu$ parameter becomes naturally around the SUSY breaking scale.
Moreover, the $E_6$ gauge singlet fields $S$ and $Z$ can be identified with 
composite operators, for example, we can take
$Z\sim \bar CC$ or $Z\sim \bar \Phi\Phi$.

In our $E_6$
case, the minimal field  contents are, in addition to $\Theta$, 
three matter multiplets, $\Psi_i({\bf 27})$, 
the pair of Higgs fields $\Phi({\bf 27})\bar \Phi({\bf 27})$ and
 the pair of Higgs fields 
$C({\bf 27})$ and $\bar C({\bf \overline{27}})$,
which are needed for the breaking  $E_6
\rightarrow SO(10)\rightarrow SU(5) \rightarrow$ the standard gauge
groups, together with  an adjoint field $A(\bf 78)$, which 
also provides a natural 
D-T splitting mechanism,  as explained in the context of the 
$SO(10)$ model  in 
separate papers,\cite{maekawa,BKM} and leaves light Higgs doublets $H$.
Among those minimal contents of matter and Higgs fields, 
we have nine charges, 
$(\psi_1,\psi_2,\psi_3), (\phi,\bar\phi), (c, \bar c), a$ and $h$, 
which determine the main features of the mass matrices of 
quarks and leptons. 
First, the CKM mixing angle almost fixes the  charges of the matter fields 
$\psi_i=(n+3,n+2,n)$, and the doublet Higgs $h=-2n$, and in order to get
bi-large mixing angles for the neutrino sector,
we need a constraint on the 
charges of the Higgs fields, $r=1/2$, i.e., $c-\bar c=\phi-\bar \phi +1$,
and also we have the constraint $\phi-\bar\phi=2n-10-l$ 
($-2\leq l \leq 2$) in order to give the proper neutrino masses. 
If we choose the charge as $a=-2$ in order
to insure that the GUT relation between the masses of down-type quarks 
and charged leptons acts only for third generation, there remain only
three degrees of freedom.
Moreover, it may be possible to build DT splitting models in which the 
light Higgs can be identified with the components of 
$\Phi$ or $C$. Actually, it is naturally realized that 
$\Phi$ can play the role of $H$, so in that case,
$\phi=h$.\cite{BKM}
This implies that there are now two degrees of freedom.
If we further impose the condition of solving the $\mu$ problem
(\ref{mu}), there remains just one degree of freedom.
It is quite suggestive that
there is a set of charge assignments that can satisfy all the above
conditions. Actually, if we take 
$n=2,\phi=h=-4,\bar \phi=0,c=-4,\bar c=-1$,
all the above conditions are satisfied for $l=-2$.
The charge assignment
$n=2, \phi=h=-4,\bar \phi=3,c=-6,\bar c=0,l=1$ is quite interesting,
because the composite operator $\bar \Phi\Phi$ can even play the same role
as the $\Theta$ field when $\xi^2\sim \lambda$; that is,
$\VEV{\bar \Phi\Phi}\sim \lambda\equiv \xi^2$.
In this case, we have the minimum model in which there are
$\Psi_1$, $\Psi_2$, $\Psi_3$, $\Phi$, $\bar \Phi$, $C$, $\bar C$ and $A$,
where all the charges are uniquely determined.

What is interesting in the $E_6$ unified model is that the condition for
suppression
of FCNC is automatically satisfied.
The essential point is that the first and second generation fields of
${\bf \bar 5}$ have the same anomalous $U(1)_A$ charge because these
fields originate from a single field $\Psi_1$.

The aspect of the family structure that has recently been 
made clear by the neutrino experiments gives 
a guide to investigate the origin of the 
family. 
The scenario discussed here is quite impressive, and 
it leads us to expect 
that we may find ``the real GUT'' in the near future.

\section*{Acknowledgements}
We would like to express our sincere thanks to T. Kugo
for his collaboration in the early stage of this work,
reading this manuscript and useful comments.
Thanks are also due to H. Nakano, K. Kurosawa, M. Yamaguchi
and S. Yamashita for instructive and interesting discussions.
We were very much stimulated by
discussions with the members
who attended to a series of research meetings [supported by
Grants-in-Aid for Scientific Research on Priority Area A 
``Neutrinos" (Y. Suzuki)].
One of the 
authors (M. B.) is supported in part by Grants-in-Aid for
Scientific Research Nos. 12047225(A2) and 12640295(C2) from
the Ministry of Education, Culture, Sports, Science and Technology
of Japan.

\appendix

\section{}
In this appendix, we explain how the vacua of the Higgs fields are
determined by the anomalous
$U(1)_A$
quantum numbers.

First, we show that none of the fields with positive anomalous
$U(1)_A$ charge acquire nonzero VEV if the Froggatt-Nielsen (FN) mechanism
acts effectively in the vacuum.
 Let the gauge singlet fields be $Z_i^\pm$ ($i=1,2,\cdots n_\pm$)
 with charges $z_i^\pm$, when $z_i^+>0$ and $z_i^-<0$.
From the $F$ flatness conditions of the superpotential,
we get $n=n_++n_-$ equations  plus one $D$-flatness condition,
\begin{equation}
 \frac{\delta  W}{\delta Z_i}=0, \qquad D_A=g_A
      \left(\sum_i z_i |Z_i|^2 +\xi^2 \right)=0,
\label{eq:fflat}
\end{equation}
where $\xi^2=\frac{g_s^2\tr Q_A}{192\pi^2} (\equiv \lambda^2 \Mp^2)$.
At first glance, these look to be over-determined. However,
 the $F$ flatness
conditions are not independent, because the gauge invariance of the
superpotential $W$ leads to the relation\footnote{
We thank H. Nakano for pointing out this relation.}
\begin{equation}
\frac{\delta  W}{\delta Z_i}z_iZ_i=0.
\label{constraint}
\end{equation}
Therefore, generically a SUSY vacuum with $\VEV{Z_i}\sim \Mp$ exists
(Vacuum a),
because the coefficients of the above conditions are generally
of order 1.
However, if $n_+\leq n_-$, we can choose another vacuum (Vacuum b)
with $\VEV{Z_i^+}=0$, which automatically satisfies the $F$-flatness
conditions $\frac{\delta  W}{\delta Z_i^-}=0$. Then the $\VEV{Z_i^-}$
are determined by the $F$-flatness conditions
$\frac{\delta  W}{\delta Z_i^+}=0$ with the constraint
(\ref{constraint}) and the $D$-flatness condition $D_A=0$.
Note that if $\lambda<1$ (i.e., $\xi<1$), 
the VEVs of $Z_i^-$ are less than the
Planck scale. This can lead to the Froggatt-Nielsen mechanism.
If we fix the normalization of $U(1)_A$ gauge symmetry so that
the largest value $z_1^-$ in the negative charges $z_i^-$ equals -1,
then the VEV of the field
$Z_1^-$ is determined from $D_A=0$ as $\VEV{Z_1^-}\sim\lambda$,
which breaks $U(1)_A$ gauge symmetry. (The field $Z_1^-$ was
introduced in the previous section as $\Theta$.) 
Other VEVs are determined by the $F$-flatness conditions of $Z_i^+$ as
$\VEV{Z_i^-}\sim \lambda^{-z_i^-}$, which is shown below.
Since $\VEV{Z_i^+}=0$, it is sufficient to examine the terms linear
in $Z_i^+$ in the superpotential in order to determine 
$\VEV{Z_i^-}$. Therefore, in general
the superpotential can be written
\begin{eqnarray}
W&=&\sum_i^{n_+}W_{Z_i^+},\\
W_{Z_i^+}&=& \lambda^{z_i^+}Z_i^+\left(\sum_j^{n_-}\lambda^{z_j^-}Z_j^-
+\sum_{j,k}^{n_-}\lambda^{z_j^-+z_k^-}Z_j^-Z_k^-+\cdots \right)\\
&=&\sum_i^{n_+} \tilde Z_i^+\left(\sum_j^{n_-}\tilde Z_j^-
+\sum_{j,k}^{n_-}\tilde Z_j^-\tilde Z_k^-+\cdots \right),
\end{eqnarray}
where $\tilde Z_i\equiv \lambda^{z_i}Z_i$.
The $F$-flatness conditions of the $Z_i^+$ fields require
\begin{equation}
\lambda^{z_i^+}\left(1+\sum_j\tilde Z_j^-+\cdots\right)=0,
\end{equation}
which generally lead to solutions $\tilde Z_j\sim O(1)$
if these $F$-flatness conditions determine the VEVs.
Thus the F-flatness condition requires
\begin{equation}
   \VEV{ Z_j} \sim O(\lambda ^{-z_j}).
\label{eq:}
\end{equation}
Here we have examined the VEVs of singlets fields, but generally
the gauge invariant operator $O$ with negative charge $o$ has
non-vanishing VEV $\VEV{O}\sim \lambda^{-o}$ if the $F$-flatness
conditions determine the VEV.

If Vacuum a is selected, the anomalous $U(1)_A$ gauge symmetry
is broken at the Planck scale, and the FN mechanism does not act.
Therefore, we cannot know the existence of the $U(1)_A$ gauge symmetry
from the low energy physics. On the other hand, if Vacuum b is
selected, the FN mechanism acts effectively, and we can understand
the signature of the $U(1)_A$ gauge symmetry from the low energy
physics. Therefore, it is natural to assume that Vacuum b is
selected in our scenario, in which the $U(1)_A$ gauge symmetry
plays an important role for the FN mechanism. This amounts to assuming
that the VEVs of the fields $Z_i^+$ vanish, which guarantees that 
the SUSY zero mechanism acts effectively.

\end{document}